\documentstyle[12pt,amssymb]{article}
\def\vk{\varkappa}
\def\h{\hbar}
\def\z{\zeta}

\makeatletter
\def\lambdabar{\protect\@lambdabar}
\def\@lambdabar{%
\relax
\bgroup
\def\@tempa{\hbox{\raise.73\ht0
\hbox to0pt{\kern.25\wd0\vrule width.5\wd0
height.1pt depth.1pt\hss}\box0}}%
\mathchoice{\setbox0\hbox{$\displaystyle\lambda$}\@tempa}%
{\setbox0\hbox{$\textstyle\lambda$}\@tempa}%
{\setbox0\hbox{$\sc00riptstyle\lambda$}\@tempa}%
{\setbox0\hbox{$\scriptscriptstyle\lambda$}\@tempa}%
\egroup
}
\makeatother

\makeatletter
\@addtoreset{equation}{section}
\makeatother

\begin{document}
\title{The quantum effects in the undulator of infinite length}
\author{V.G.Bagrov$^1$, V.V.Belov$^2$, M.M. Nikitin$^3$, and
A.Yu.Trifonov$^3$}
\date{$^1${\it High Current Electronics Institute,\\
Siberian Division Russian Academy of Science\\
4 Akademichesky Ave., 634055 Tomsk Russia\\
$^2$Department of Applied Mathematics\\
Moscow Institute of Electronic and Mathematics,\\
B. Vusovsky 3/12, 109028 Moscow, Russia\\
$^3$Department of Mathematical Physics,\\
Tomsk Polytechnical University, 30 Lenin Ave., 634034 Tomsk, Russia}}
\maketitle

\abstract
The first order quantum correction to the power of spontaneous
radiation of electrons in an arbitrary two-component periodic magnetic
field was obtained. The phenomenon of selfpolarization of the spin of
electrons in a process of spontaneous radiation was also studied. By
electron's motion in a spiral magnetic undulator, the quantitative
characteristics of selfpolarization (the polarization degree and the
relaxation time) are different from corresponding ones in synchrotron
radiation. The limiting cases of near-axis and ultrarelativistic
approximation were considered.

\section*{Introduction}

The investigation of spontaneous radiation of electrons moving in
periodic structures (undulators) is now an important and rather well
developed branch of modern physics. A detailed bibliography about this
problem can be found, for example, in reviews [1--2] and monographs [3--5].

In the majority of studies, the radiation was investigated by
methods of classical electrodynamics [1--5]. Quantum-electrodynamical
considerations were applied in few studies [1, 6--15],
where the character of quantum corrections to the radiation of
ultrarelativistic particles ($\gamma\gg 1$) was calculated and, in
particular, it was shown that these corrections are small in actual
undulators.

Nevertheless, because of wide usage in theoretical and experimental
research of magnetic undulators, it is important to give a detailed
analysis of the character of quantum corrections to the radiation
over the whole range of electron energy allowed by the undulator regime.
This range also includes the nonrelativistic energies, for which
the effects caused by particles motion are significant.

In the present paper we considered the first quantum correction to
the characteristic of undulator radiation of charged spinor particles
in a magnetic undulator. This consideration is limited to the case
of a vector potential of an arbitrary magnetic field depending only
on one spatial coordinate (along the particle drift). This particular
model of an external magnetic field allows us to consider typical
regimes of partical motion in magnetic undulators: along the plane
periodic trajectory for a plane undulator, and along the helical
trajectory for a spiral undulator. On the other hand, this model
allowes us effective calculate quantum-electrodynamical characteristics
by using Bloch's stationary wave functions obtained in [16] by
the one-dimensional WKB method [17].

Let us give some general conclusions. The results of our
calculations [9--15] which are of certain theoretical interest for the
problems of motion and radiation of particles in magnetic undulators of all
energies when the quantum (spinor) properties are taken into account.

1. If the characteristics of spontaneous radiation of electrons in a
periodic magnetic field (calculated by methods of quantum theory in the
first order over the radiation field when the motion is
semiclassical) are represented in the form of formal (and generally
asymptotic) Taylor series in the Planck constant $\h$ as $\h
\to 0$, then each term of this series can be represented as a
functional of the classical particle trajectory, the external field, and
derivatives of this field on the classical trajectory. In particular, the
$n$ quantum correction contains the $n+2$-derivative of velocity,
i.e., $n+1$-derivative of the vector potential. In the case of $n=1$
[18], we calculated boson radiation in an arbitrary two-component periodical
magnetic field, in which the helical motion is realized.

It should be mentioned that the dependence of the characteristics of
spontaneous radiation (calculated in the semi-classical
approximation as $\h\to 0$ with relativistic accuracy to the first order
of $\h$) on the parameters of classical trajectory was
first shown in [19--20] for the case of electron energies in the
relativistic range.

2. The quantum corrections to the radiation characteristics and the influence
of different quantum effects (in the whole energy
range allowed by the undulator regime of electron motion) depend\footnote{In
the ultrarelativistic case ($\gamma\to\infty$), the dependence on the
structure of an external field can be neglected.} on the specific form of
the potential of the external field. For electrons moving along the same classical
trajectory in different external fields, the quantum corrections (in
comparison with the classical term) in general will differ.

This conclusion is illustrated here by calculations of
the characteristics of radiation of spinor particles moving in magnetic
undulators. For example:

1) The explicit dependence of quantum corrections to the frequency of the
photon radiation on the specific form of potential of the external
periodical magnetic field is shown (see Eq. (2.9)).

2) The quantum expansion parameter of the total power of undulator radiation
(in the on-axis approximation) qualitatively differs from the corresponding
parameter in synchrotron radiation and depends explicitly on the character
of magnetic field and its first derivative on the classical trajectory.
It is interesting that, for a specific form of the field in which the particle
moves along the spiral trajectory, in the ultrarelativistic limit the first
quantum correction to the full radiation power coincides with the
expression of the quantum term in the power of synchrotron radiation of the
charge moving along the spiral in a constant and homogeneous magnetic
field [21].

3) The quantitative characteristics of the effect of radiational
self-polarization of electrons in a helical undulator (the degree of
polarization and the relaxation time) differ from those in the theory of
synchrotron radiation [22] and also from similar results in a field of
a flat electromagnetic wave [23] and in a axisymmetric focusing electric
field [24], though in all considered (examined) cases a electron movement
on a spiral is realized. This fact proves that the structure of the
external field but not the classical electron trajectory plays a
significant role in the phenomena of self-polarization.
It should be mentioned that, in the ultrarelativistic limit
($\gamma\to 0$), the characteristics of the process of
self-polarization calculated in Sec. 6 coincide (with an accuracy of
$\sim\gamma^6$) with those obtained earlier in [7] on the basis of an
operator semi-classical method of calculation.

\section{Semi-classical electron wave function }

Let us consider an electron moving in a stationary magnetic field
with the  vector potential :
\begin{equation}
\begin{array}{c}
\vec{\cal A}(z)=({\cal A}_1(z),{\cal A}_2(z),0),\\
\vec{\cal A}(z+l)=\vec{\cal A}(z),
\end{array}
\end{equation}
where ${\cal A}_i(z)$ are arbitrary smooth periodic functions, $l$
is the field period. Let us describe the electron motion on the basis of the
relativistic Dirac equation:
\begin{eqnarray}
& \hat{\cal H}_D\Psi=E\Psi,\qquad
\hat{\cal H}_D=c(\vec{\alpha},\hat{\vec{\cal P}})+\rho_3m_0c^2,\\
\nonumber 0 & \hat{\vec{\cal P}}=(-i\h\nabla)-\displaystyle\frac
ec\vec{\cal A}(z),\\ & \vec{\alpha}=\left(\begin{array}{ll} 0
&\vec{\sigma}\\ \vec{\sigma}&0\end{array}\right),\qquad
\rho_3=\left(\begin{array}{ll}
1&\;0\\
0&-1\end{array}\right),\nonumber
\end{eqnarray}
where $\vec\sigma=(\sigma_1,\sigma_2,\sigma_3)$ are Pauli matrices,
$E$ is the electron energy and $e=-e_0$ is the electron charge. Let
us separate the solutions of equation (1.2) over a polarization states by
condition:
\begin{eqnarray}{c}
& \hat S_t\Psi=\z\lambda\Psi, \qquad
\hat S_t=(\vec\Sigma',\hat{\vec{\cal P}}),\\ \nonumber 0
& \z=\pm1,\;\;\;\lambda^2=c^{-2}(E^2-m_0^2c^4),\\
& \vec\Sigma'=\left(\begin{array}{ll}
\vec{\sigma}&0\\
0           &\vec{\sigma}
\end{array}\right).\nonumber
\end{eqnarray}
Here the parameter $\z$ describes the spin orientation with
reference to the direction of motion (longitudinal polarization) either
along the direction of motion ($\z=+1$) or in the opposite direction
($\z=-1$) [25].

Taking into account of quantum integrals of motion
$$
\hat p_1\Psi=p_1\Psi,\qquad
\hat p_2\Psi=p_2\Psi,\qquad
\hat{\vec p}=-i\h\nabla,
$$
the solution of system (1.2) and (1.3) will be
represented by:
\begin{equation}
\Psi_{E,\vec p_\perp,\z}=N \exp
\Big\{\frac{i}{\h}(\vec{p}_{\perp},\vec{r})\Big\}
\phi_{E,\vec p_\perp,\z}(z,\h),
\end{equation}
where
\begin{equation}
\begin{array}{c}
\phi_{E,\vec p_\perp,\z}(z,\h)=(g_{E,\vec p_\perp,\z}(z,\h),
\chi_{E,\vec p_\perp,\z}(z,\h))^t,\;\;\;
\vec p_{\perp}=(p_1,p_2,0),\\
\chi_{E,\vec p_\perp,\z}(z,\h)=\displaystyle\frac{\z c\lambda}
{E+m_0c^2} g_{E,\vec p_\perp,\z}(z,\h),\;\;\;
\vec r=(x,y,z),\end{array}
\end{equation}
$N$ is a normalization constant. Spinor $g_{E,\vec p_\perp,\z}(z,\h)$
satisfies the condition
\begin{eqnarray}
& \left\{(\vec{\sigma},\vec{\cal P}_{\perp})+\sigma_3\hat p_3\right\}
g_{E,\vec p_\perp,\z}(z,\h)=\z\lambda g_{E,\vec p_\perp,\z}(z,\h),\\ \nonumber
& \vec{\cal P}_{\perp}=({\cal P}_1,{\cal P}_2,0),\\
& {\cal P}_i=p_i+\displaystyle\frac{e_0}{c}{\cal A}_i(z),\;\;\;
i=1,2\nonumber
\end{eqnarray}
or in the equivalent form $g_{E,\vec p_\perp,\z}(z,\h)=
(g_+(z,\h,E,\vec p_\perp,\z),g_-(z,\h,E,\vec p_\perp,\z))^t$
\footnote{In case this couses no misanderstanding the indeces
$z,\h,E,\vec p_\perp,\z$ may be omitted.}
\begin{equation}
\left\{
\begin{array}{ll}
\{(\hat p_3+\z\lambda)({\cal P}_1-i{\cal P}_2)^{-1}
(\hat p_3-\z\lambda)+{\cal P}_1+i{\cal P}_2\}g_+(z,\h)=0,\\
g_{-}(z,\h)=({\cal P}_1-i{\cal P}_2)^{-1}(\z\lambda-
\hat p_3)g_+(z,\h).\end{array}\right.
\end{equation}
In the absence of turning points
\begin{equation}
p^2(z)=\lambda^2-(\vec{\cal P}_{\perp},\vec{\cal P}_{\perp})>0
\end{equation}
the WKB-solution of system (1.6) has the form [26]:
\begin{equation}
g_+(z,\h)=f_+(z,\h)\exp\Big\{\frac{i}{\h}S(z)\Big\},
\end{equation}
where $f_+(z,\h)=f_+^{(0)}(z,\h)+\h f_+^{(1)}(z,\h)$ is a
regular series over $ \h\to 0$. Here, the actual measureless
expansion parameter is the ratio of the de'Brojlie electron wave length to
the period of changing of field $\lambdabar\approx\frac{\h\omega_0}
{E\beta^2_{\parallel}}$, where [19]
\[ T\omega_0=2\pi, \qquad c\beta_{\parallel}T=l, \qquad
T=\frac{E}{c^2} \int ^l_0\frac{dz}{p(z)}. \]
Substituting (1.9) into (1.7) we obtain:
\begin{eqnarray}
& S(z)=\displaystyle\int \limits^z_0p(z)dz,\\ \nonumber
& f_+^{(0)}(z)=\exp\Big\{-\displaystyle\int\limits^{z}_0dz\Big[
\displaystyle\frac{({\cal P}_1-i{\cal P}_2)}
{2p(z)}\frac{d}{dz}\frac{(p(z)-\z\lambda)}{({\cal P}_1-i
{\cal P}_2)}\Big]\Big\},\\
& f_+^{(1)}(z)=if_+^{(0)}(z) \displaystyle\int\limits^z_0dz
\Big[\frac{({\cal P}_1+i{\cal P}_2)}
{2p(z)f_+^{(0)}(z)}\frac{d}{dz}\frac{\dot f_+^{(0)}(z)}{({\cal P}_1-
i{\cal P}_2)}\Big].\nonumber
\end{eqnarray}
Here, the dot defines the differentiation with respect to the parameter
$z$ $(\dot{\z}=\frac{d\z}{dz})$. By condition (1.8), the spinor
$g(z,\h)=(g_+(z,\h),g_-(z,\h))^t$, where
\begin{eqnarray}
& g_\pm(z,\h)=\mu_\pm\Big\{\displaystyle\frac{({\cal P}_1\mp
i{\cal P}_2) (\lambda\mp\z p)}
{(\vec{\cal P}_{\perp},\vec{\cal P}_{\perp})^{1/2}p}
\Big\}^{1/2}\exp\Big\{\frac{i}{\h}\int \limits^{z}_0
\Big(p+\h\frac{\z\lambda}{p}
\frac{({\dot{\cal P}}_2{\cal P}_1-{\dot{\cal P}}_1{\cal P}_2)}
{(\vec{\cal P}_{\perp},\vec{\cal P}_{\perp})}\Big)dz\Big\}\times\\ \nonumber
&\times\left\{ 1-\displaystyle\frac{i\h}{4}\left( \frac{\dot p}{p^2}-
\frac{\dot{\cal P}_1\mp i\dot{\cal P}_2}{{\cal P}_1\mp
i{\cal P}_2}\frac{p\mp\z\lambda}{p^2}+\int \limits^{z}_0
\frac{(\dot{\cal P}_2{\cal P}_1-\dot{\cal P}_1{\cal P}_2)^2+
\lambda^2\dot p^2}
{2p^3(\vec{\cal P}_{\perp},\vec{\cal P}_{\perp})}dz\right)\right\},\\
& \mu_+=1,\;\;\; \mu_{-}=\z,\;\;\; p=p(z)\nonumber
\end{eqnarray}
satisfies (1.6) with accuracy up to $O(\h^2)$. From here and from the
condition
$$
g_{E,\vec p_\perp,\z}(z+l,\h)=\exp\left\{\frac{i}{\h}
q(E,\vec p_\perp,\z)l\right\}g_{E,\vec p_\perp,\z}(z,\h)
$$
we find the quasi-momentum of an electron:
\begin{equation}
q(E,\vec p_{\perp},\z)=\frac{1}{l}\int\limits^{l}_0
\Big\{p+\z\frac{\h\lambda}{2p}\frac{\dot{\cal P}_2{\cal P}_1-
\dot{\cal P}_1{\cal P}_2}{(\vec{\cal P}_{\perp},\vec{\cal P}_{\perp})}-
\frac{\h^2}{8p^3}\frac{(\dot{\cal P}_2{\cal P}_1-\dot{\cal P}_1
{\cal P}_2)^2+\lambda^2{\dot{p}}^2}{(\vec{\cal P}_{\perp},
\vec{\cal P}_{\perp})}\Big\}dz.
\end{equation}
In the absence of turning points $p^2(z)>0$, the normalization condition on
the electron charge for the semiclassical stationary solutions (1.4)
of the Dirac equation
$$
\int \Psi^{+}_{b'}\Psi_{b}d^{3}x=\delta_{b',b},
\;\;\;
b=(E,\vec{p}_{\perp},\z)
$$
is  equivalent to the condition
$$
4L^2NN'\Big[1+\frac{\z\z'c^2\lambda\lambda'}{(E+m_0c^2)
(E'+m_0c^2)}\Big]\int\limits^{L}_{-L}dz\,g^{+}_{E',\vec{p}_{\perp},\z}
g_{E,\vec p_{\perp},\z'}=\delta_{E',E}\delta_{\z,\z'}.
$$
By analogy with the case without spin [12], we obtain
\begin{eqnarray*}
&\displaystyle\int\limits^L_{-L}g^{+}_{E',\vec{p}_{\perp},\z}(z)
g_{E,\vec{p}_{\perp},\z}(z)dz=2L\delta_{qq'}\frac{1}{l}
\int\limits^l_0dzg^{+}_{E,\vec{p}_{\perp},\z}g_{E,\vec{p}_{\perp},\z},\\
& \displaystyle\delta_{qq'}=\left\{\Big|\frac{\partial(q-q')}{\partial(E-E')}
\Big|_{q-q'=0}\right\}^{-1}\delta_{E,E'},\\
& g^+_{E,\vec{p}_{\perp},\z}g_{E,\vec{p}_{\perp},-\z}=O(\h).
\end{eqnarray*}
Here, we used the fact that the quasi-momentum is defined with the
accuracy up to $\frac{2\pi\h}l n$:
$$
\begin{array}{c}
\displaystyle\int\limits_{-L}^Ldz\,4L^2N^2g^{+}_{E',\vec{p}_{\perp},\z'}
(z,\h)g_{E,\vec p_{\perp},\z}(z,\h)
\left(1+\frac{\z\z'c^2\lambda\lambda'}
{(E+m_0c^2)(E'+m_0c^2)}\right)=\\
=4L^2N^2\displaystyle\frac{2E}{E+m_0c^2}\left\{\frac 1l
\int\limits^l_0 dz\,g^+_{E,\vec p_{\perp},\z}
g_{E,\vec p_{\perp},\z}\right\}
\left\{\left|
\frac{\partial(q-q')}{\partial(E-E')}
\right|_{q-q'=0}\right\}^{-1}
\delta_{E,E'}\delta_{\z,\z'},\\
\left|\displaystyle
\frac{\partial(q-q')}{\partial(E-E')}\right|_{q-q'=0}=
\frac{E}{c^2l}\int \limits^l_0\frac{dz}{p}
\left\{1-\z\frac{\h}{2}
\frac{{\cal P}_1\dot{\cal P}_2-\dot{\cal P}_1{\cal P}_2}
{\lambda p^2}\right\} ,\\
\displaystyle\frac 1l \int\limits^l_0 g^+_{E,\vec p_{\perp},\z}(z)
g_{E,\vec p_{\perp},\z}(z)dz=\frac 1l \int \limits^l_0\frac{dz}{p}
\left\{
2\lambda-\z\h
\frac{\dot{\cal P}_1{\cal P}_2-\dot{\cal P}_1{\cal P}_2}{p^2}
\right\}.
\end{array}
$$
As a result, for the normalizing factor we obtain
$$
N^2=\frac{E+m_0c^2}{4c\lambda(2L)^{3}}.
$$

\section{The spectral-angular distribution of power}

The power radiated to the element of solid angle
$d\Omega=\sin{\theta}d\theta d\varphi$ at the electronic transition from the
state $\Psi_b$ to the state $\Psi_{b'}$ with the radiation of
photon can be found by usual electrodynamical methods and is equal to
[21]:
\begin{eqnarray}
& \begin{array}{c}
\displaystyle\frac{dW_{b,b'}}{d\Omega}=\frac{ce^2_0}{2\pi}
\int \limits_0^{\infty}d\vk\,\vk^2\delta\Big(\vk-\frac{E-E'}{c\h}\Big)
\{|\alpha_{\pi}|^2+|\alpha_{\sigma}|^2\},\\
\alpha_{\pi}=(\vec e_\pi,\vec B_{b,b'}),
\;\;\;
\alpha_\sigma=(\vec e_\sigma,\vec B_{b,b'});
\end{array}\\
& \begin{array}{c}
\vec B_{b,b'}=\displaystyle\int
\Psi_{b'}^+\vec\alpha\Psi_b\exp\{-i\vk(\vec n,\vec r)\}d\vec x,\\
\vec e_\pi=(\cos\varphi\cos\theta,\sin\varphi\cos\theta,-\sin\theta),\\
\vec e_\sigma=(\sin\varphi,-\cos\varphi,0),\\
\vec n=(e_1,e_2,e_3),\\
e_1=\cos\varphi\sin\theta,
\;\;\;
e_2=\sin\varphi\sin\theta,
\;\;\;
e_3=\cos\theta,
\end{array}
\end{eqnarray}
where $\vec\alpha$ are the Dirac matrices and $\h\vk\vec{n}$
is the momentum  of the radiated photon. We assumed,
that the average electron drift along the axes $Ox$ and $Oy$ is absent:
\begin{equation}
\int \limits^{l}_0\frac{{\cal P}_1(z)}{p(z)}dz=
\int \limits^{l}_0\frac{{\cal P}_2(z)}{p(z)}dz=0.
\end{equation}

The spectral-angular distribution  of the total radiation power can be
obtained from (2.1) by summation over all final states
$E',\vec p{}'_{\perp}$:
$$
\frac{dW(\z,\z')}{d\Omega}=\sum _{E',\vec p{}'_{\perp}}
\frac{dW_{b,b'}}{d\Omega}.
$$

By calculating the matrix elements (2.2) the wave functions (1.4) are
represented in a form of Bloch functions and their periodical parts are
expanded in Fourier series. After integration over coordinates, this
leads to the following conservation laws:
\begin{equation}
\left\{\begin{array}{l}
\vec{p}_{\perp}-\vec p{}'_{\perp}=\h\vec{e}_{\perp}\vk,\\
q-q'=e_3\h\vk+2\pi\h n/l.
\end{array}\right.
\end{equation}

For the spectral-angular distribution of radiation power we obtain:
\begin{eqnarray}
&\displaystyle\frac{dW(\z,\z')}{d\Omega}=\frac{c\beta_{\parallel}e_0^2}{2\pi}
\sum^{\infty}_{n=1}\vk^2\Big/{\left|\frac{\partial\Phi(\vk)}{\partial
\vk}\right|_{\Phi(\vk)=0}}
\left\{|(\vec e_\pi,\vec B(n,\z,\z'))|^2+
|(\vec e_\sigma,\vec B(n,\z,\z'))|^2\right\};\\
& \vec B(n,\z,\z')=(2L)^{3}NN'\displaystyle\frac 1l
\int\limits_0^l dz\,e^{-i\vk e_3z}\phi^+_{E',\vec{p}'_{\perp},\z'}(z)
\vec\alpha\phi_{E,\vec{p}_{\perp},\z}(z),\\
& \Phi(\vk)=\h^{-1}(q-q')-\vk e_3-2\pi n/l.\nonumber
\end{eqnarray}

The radiation frequency $\omega(n,\h)$ can be defined from the
system of  conservation laws (2.4) and the conservation law
of energy $E-E'=c\h\vk$. For this, we find the increment the
of quasi-momentum $\Delta q=q-q'$, which, according to (1.11), is equal
to
\begin{equation}
\Delta q=\h\omega\rho_0(l)+
\frac{1}{2}(\z-\z')\h\rho_3(l)+
\frac{1}{2}\h^2\omega^2\rho_1(l)+
\frac{1}{2}\h^2\omega\z'\rho_2(l),
\end{equation}
where
\begin{eqnarray}
&& \rho_0(x)=\displaystyle\frac{1}{cx} \int \limits^x_0
\left(\frac{E}{c}-
(\vec{e}_{\perp},\vec{\cal P}_{\perp})\right)\frac{dz}{p(z)},\\ \nonumber
&&\rho_1(x)=\displaystyle\frac{1}{c^2x}
\int \limits^x_0\left\{\left(\frac{E}{c}-
(\vec{e}_{\perp},\vec{\cal P}_{\perp}
)\right)^2-p^2e_3^2\right\}\frac{dz}{p^3(z)},\\ \nonumber
&&\rho_2(x)=\displaystyle\frac{\lambda}{cx}
\int \limits^x_0
\left\{
\frac{ {\cal P}_2e_1-{\cal P}_1e_2 }
{ {\cal P}_1^2+{\cal P}_2^2 }-
\frac{
\dot{\cal P}_2{\cal P}_1-\dot{\cal P}_1{\cal P}_2 }
{ {\cal P}_1^2+{\cal P}_2^2 }
\right.\times \\
&&\qquad \times
\left.
\left(
\displaystyle\frac{E}{c^2}
\frac{(\vec{\cal P}_{\perp},\vec{\cal P}_{\perp})}{\lambda^2p^2}
-\frac{({\vec{e}}_{\perp},\vec{\cal P}_{\perp})}
{({\cal P}_1^2+{\cal P}_2^2)p^2}
\right)
\left(
(\vec{\cal P}_{\perp},\vec{\cal P}_{\perp})-2p^2)
\right)
\right\}
\frac{dz}{p(z)},\\ \nonumber
&& \rho_3(x)=\displaystyle\frac\lambda x\int\limits^x_0
\frac{\dot{\cal P}_2{\cal P}_1-\dot{\cal P}_1{\cal P}_2}
{(\vec{\cal P}_{\perp},\vec{\cal P}_{\perp})p}dz.\nonumber
\end{eqnarray}

Taking into account the condition (2.3), we will get
$\rho_0(l)=1/(c\beta_{\parallel})$. Substituting (2.7) into (2.4), we find
the radiation frequency (up to $O(\h^2)$) in the form:
\begin{eqnarray}
& \omega(n,\h)=\omega_{\rm cl}\displaystyle\left\{
1-\frac \h2\frac{c\beta_{\parallel}}{\psi_0}
(\omega_{\rm cl}\rho_1(l)+\z'\rho_2(l))\right\};\\
& \omega_{\rm cl}=\displaystyle\frac{\beta_{\parallel}c}{\psi_0}\left\{
\omega_{n}-\frac 12(\z-\z')\rho_3(l)\right\},\\
& \psi_0=1-\beta_{\parallel}e_3,
\qquad \omega_{n}=2\pi n/l.\nonumber
\end{eqnarray}

By analogy with (2.7), we obtain
\begin{equation}
\left|\frac{\partial\Phi}{\partial\vk}\right|_{\Phi(\vk)
=0}=
\frac{\psi_0}{\beta_{\parallel}}
\left\{
1+\frac{\h c\beta_{\parallel}}{\psi_0}
(\omega\rho_1(l)+\frac{\z}{2}\rho_2(l))
\right\}.
\end{equation}

It should be noted, that the expression for the frequency of
classical radiation (2.9) thus obtained differs from the corresponding
expression in [12] by a summand, which is proportional to
$(\z-\z')\rho_3(l)$ and corresponds to transitions with
spin flip. The probability of these transitions is proportional to
$\h$. Therefore, at $\h=0$, the frequency (1.9) differs from the
frequency of classical radiation in the field (1.1) to harmonics whose the
probability of radiation is equal to zero. This means
that they just coincide.

\section{The matrix element of the transition currents}

Substituting (1.5) into (2.6), we obtain
\begin{eqnarray}
&& \vec B(n,\z,\z')=\displaystyle\frac 12(1+\z\z')\vec B^{\uparrow\uparrow}(n,\z)
+\frac 12(1-\z\z')\vec B^{\uparrow\downarrow}(n,\z),\\
&& \vec B^{\uparrow\uparrow}(n,\z)=\displaystyle\frac{\z}{2l}
\int \limits^{l}_0dz \exp\{-i\vk e_3z\}g^+_{E',\vec p'_\perp,\z}(z)
\vec\sigma g_{E,\vec p_\perp,\z}(z),\\
&& \vec B^{\uparrow\downarrow}(n,\z)=\displaystyle
\frac{\h\vk}{4}\frac{m_0c}{\lambda^2}\frac{\z}{l}
\int \limits^{l}_0dz \exp\{-i\vk e_3z\}g^+_{E',\vec p'_\perp,-\z}(z)
\vec{\sigma}g_{E,\vec p_\perp,\z}(z).
\end{eqnarray}

Let us consider the matrix elements without spin flip
\begin{eqnarray*}
&& \left\{
\begin{array}{ll}
g_{\pm}^*(z,E',\vec p'_\perp,\z) &g_{\pm}(z,E,\vec p_\perp,\z)\\
g_{\pm}^*(z,E',\vec p'_\perp,\z)& g_{\mp}(z,E,\vec p_\perp,\z)
\end{array}
\right\}=\\
&&\qquad=\left\{
\begin{array}{l}
f_{\pm}^{(0)}(z,E',\vec p'_\perp,\z)f_{\pm}^{(0)}(z,E,\vec p_\perp,\z)
\left( 1+\displaystyle\frac{\h}{2}
\frac{\dot{\cal P}_2{\cal P}_1-\dot{\cal P}_1{\cal P}_2}
{({\cal P}_1^2+{\cal P}_2^2)p}(p\mp\z\lambda)\right)\\
f_{\pm}^{(0)*}(z,E',\vec p'_\perp,\z)f^{(0)}_{\mp}(z,E,\vec p_\perp,\z)
\left(1\mp i\z\h\displaystyle \frac{\lambda}{p^2}
\frac{\dot{\cal P}_1\pm i\dot{\cal P}_2}
{{\cal P}_1\pm i{\cal P}_2}\right)
\end{array}
\right\}\times\\
&& \qquad\times \exp\left\{
\displaystyle\frac{i}{\h}
\int\limits^z_0dz\,[p(z,E,\vec p_\perp)-p(z,E',\vec p'_\perp)]
\right\};\\
&& f_{\pm}^{(0)*}(z,E',\vec p'_\perp,\z)f_{\pm}^{(0)}(z,E,\vec p_\perp,\z)=
\left[\displaystyle\frac{\lambda+\z p}{p}-
\frac{\h\vk}{2p}
\left\{
\frac{\lambda\pm\z p}{p^2}\Big[({\vec{e}}_{\perp},\vec{\cal P}_{\perp})-
\frac{E}{c}\Big]\pm
\right.\right.\\
&& \qquad\left.\left.
\pm i\displaystyle\frac{e_1{\cal P}_2-e_2{\cal P}_1}
{{\cal P}_1^2+{\cal P}_2^2}\pm\z
\frac{
E/c-(\vec e_{\perp},\vec{\cal P}_{\perp})}p+
\frac E{c\lambda}\right\}\right]\times\\
&& \qquad\times
\exp\{i\z\h\vk z\rho_2(z)/2\};\\
&& f_{\pm}^{(0)}(z,E',\vec p'_\perp,\z)f_{\mp}^{(0)}(z,E,\vec p_\perp,\z)=
\displaystyle\frac\z p
\left[
{\cal P}_1\pm i{\cal P}_2-
\frac 12\h\vk
\Big\{
e_1\pm ie_2+({\cal P}_1\pm i{\cal P}_2)
\right.\times\\
&& \qquad\times
\left.
\displaystyle\frac 1{p^2}
\left(
\pm\z\lambda p
\left(
\frac E{c\lambda}-
\frac{({\vec e}_{\perp},\vec{\cal P}_{\perp})}
{{\cal P}_1^2+{\cal P}_2^2}
\right)
-\left(
\frac Ec-({\vec{e}}_{\perp},\vec{\cal P}_{\perp})
\right)
\right)
\Big\}\right]\times\\
&&\qquad\times
\exp
\{i\z\h\vk z\rho_2(z)/2\}
\end{eqnarray*}
and by analogy with the scalar case, we have
\begin{equation}
\int\limits^z_0[p(z,E,\vec p_\perp)-p(z,E',\vec p'_\perp)]dz=
\omega z\rho_0(z)+\frac 12 \h\omega^2 z\rho_1(z).
\end{equation}

So, within an accuracy up to the quantum correction of the first order over
$\h\to 0$, we obtain:
\begin{eqnarray}
&& B^{\uparrow\uparrow}_1(n,\z)=
\displaystyle\frac 1l
\int \limits^{l}_0 \frac{dz}p
e^{i\psi_1(z,n,\z,\h)}
\left[
{\cal P}_1+
\frac{\h\vk}{2}
\left\{
{\cal P}_1
\frac{E/c-(\vec e_{\perp},\vec{\cal P}_{\perp})}
{p^2}-
\right.\right.\\ \nonumber
&&\qquad \left.\left.
-e_1-\z\lambda
\displaystyle\frac{i{\cal P}_2}p
\left(
\frac E{c\lambda^2}-
\frac{(\vec e_{\perp},\vec{\cal P}_{\perp}) }
{ {\cal P}_1^2+{\cal P}_2^2 }
\right)+
\z\lambda\frac{\dot{\cal P}_2}{\vk p^2}
\right\}\right],\\
&& B_2^{\uparrow\uparrow}(n,\z)=
\displaystyle\frac 1l
\int \limits^l_0\frac{dz}p e^{i\psi_1(z,n,\z,\h)}
\left[
{\cal P}_2+\frac{\h\vk}{2}
\left\{
{\cal P}_2
\frac{E/c-(\vec e_{\perp},\vec{\cal P}_{\perp})}{p^2}-e_2-
\right.\right.\\ \nonumber
&&\qquad\left.\left.
-\z\lambda\displaystyle\frac{i{\cal P}_1}{p}
\left(
\frac E{c\lambda^2}-
\frac{(\vec e_{\perp},\vec{\cal P}_{\perp})}
{ {\cal P}_1^2+{\cal P}_2^2 }
\right)+
\z\lambda\frac{\dot{\cal P}_1}{\vk p^2}
\right\}\right],\\
&& B_3^{\uparrow\uparrow}(n,\z)=
\displaystyle\frac 1l
\int \limits^l_0\frac{dz}{p}e^{i\psi_1(z,n,\z,\h)}
\left[
p+\frac{\h\vk}{2}
\left\{
\frac{E/c-(\vec e_{\perp},\vec{\cal P}_{\perp})}p
+\right.\right.\\ \nonumber
&&\qquad\left.\left.
+i\displaystyle\frac{\dot p}{\vk p}-e_3+
i\z\lambda
\frac{ e_1{\cal P}_2-e_2{\cal P}_1 }
{ {\cal P}_1^2+{\cal P}_2^2 }
\right\}\right],
\end{eqnarray}
where
\begin{equation}
\psi_1(z,n,\z,\h)=\omega z\rho_0(z)-
\omega z\frac{e_3}{c}+
\frac{\h}{2}\omega^2z\rho_1(z)+
\z\frac{\h}{2}\omega z\rho_2(z)
\end{equation}
and $\rho_i(z)$ are defined in (2.7). The matrix elements (3.4)--(3.6)
have the structure
\begin{equation}
{\vec{B}}^{\uparrow\uparrow}(n,\z)=
\tilde{\vec B}^{\uparrow\uparrow}(n,\z)-
\frac{\h\vk}{2}\vec{n}\frac{1}{l}
\int \limits^{l}_0
\frac{dz}{p}\exp
\left\{
i\omega z
\left(
\rho_0(z)-\frac{e_3}{c}
\right)\right\}.
\end{equation}
The last term does not influence the radiation power. So we will consider
$\vec B^{\uparrow\uparrow}(n,\z)=\tilde{\vec B}^{\uparrow\uparrow}(n,\z)$.
For the expressions with the spin flip, we perform analogous
calculations:
\begin{eqnarray*}
&& f_{\pm}^{(0)*}(z,E',\vec p{}'_{\perp},-\z)f_{\pm}^{(0)}(z,E,\vec p_{\perp},\z)=
\pm({\vec{\cal P}}_{\perp},\vec{\cal P}_{\perp})^{1/2}
\exp\{i(\psi_2(z,\z)-\vk e_3z)\},\\
&& f_{\pm}^{(0)*}(z,E',\vec p{}'_{\perp},-\z)f_{\mp}^{(0)}(z,E,\vec p_{\perp},\z)=
\pm\z
\displaystyle\frac{({\cal P}_1\mp i{\cal P}_2) (\lambda\pm\z p) }
{ ({\cal P}_1^2+{\cal P}_2^2)^{1/2}}
\exp\{i(\psi_2(z,\z)-\vk z e_3) \},
\end{eqnarray*}
where
\begin{equation}
\psi_2(z,\z)=
\omega\Big(z\rho_0(z)-\frac{e_3}{c}z\Big)+
\z z\rho_3(z)=
\omega F_0(z)+\z z\rho_3(z).
\end{equation}
We obtain:
\begin{eqnarray}
&& B_1^{\uparrow\downarrow}(n,\z)=
\displaystyle\frac{\h\vk}{2}
\frac{m_0c}{\lambda^2}
\frac{1}{l}\int \limits^l_0
\frac{ i{\cal P}_2\lambda-\z{\cal P}_1p }
{ p({\cal P}_1^2+{\cal P}_2^2)^{1/2}}
\exp\{i\psi_2(z,\z)\}dz,\\
&& B_2^{\uparrow\downarrow}(n,\z)=
\displaystyle\frac{\h\vk}{2}
\frac{m_0c}{\lambda^2}\frac{1}{l}
\int \limits^l_0
\frac{ -i\lambda{\cal P}_1-\z p{\cal P}_2 }
{ p({\cal P}_1^2+{\cal P}_2^2)^{1/2}}
\exp\{i\psi_2(z,\z)\}dz,\\
&& B_3^{\uparrow\downarrow}(n,\z)=
\displaystyle\frac{\h\vk}{2}\frac{m_0c}{\lambda^2}\frac{1}{l}
\int \limits^l_0
\z\frac{({\cal P}_1^2+{\cal P}_2^2)^{1/2}}p
\exp\{i\psi_2(z,\z)\}dz.
\end{eqnarray}

The obtained expressions (2.5), (2.8), (2.10), (3.1)--(3.11), in
principle solve the problem of taking into account the
quantum corrections caused by quantum effects of motion
itself $\h\omega_0/(E\beta_{\parallel})$, and by output of radiated
photon $\h\omega/E$.

\section{The full radiation power in the on-axis approximation}

To integrate the spectral-angular distribution over all angles and to
sum over the spectrum in order to obtain the total radiation power with
an accuracy to the first quantum correction is possible only in the near-axis
approximation. It can be characterized by the classical parameter
\begin{equation}
\mu=
\mathop{\rm max}\limits_{t\in[0,T]}
\left\{
\left|
\frac{\beta_1}{\beta_3}
\right|,\;
\left|
\frac{\beta_2}{\beta_3}
\right|
\right\}
\end{equation}
which is the maximum deflection angle of the electron velocity from the
undulator axis. Here $c\beta_j=c^2{\cal P}_j/E$ is the velocity along
the $j$ is the axis, $T$ is the period of motion over trajectories
$$
T=\frac{E}{c^2}\int\limits^{l}_0\frac{dz}{p(z)}
$$
which is related to the period of field $l$ by
$l=c\beta_{\parallel}T$, where $c\beta_{\parallel}=[\rho_1(l)]^{-1}$ is an
average drift along the axis $Oz$, and
$\vec{\cal P}=({\cal P}_1,{\cal P}_2,p)$ is a kinetic momentum.

Summing the spectral-angular distribution of the radiation power over a
spin and integrating over the time in (2.8), (3.1) and (3.4)--(3.6),
we use the following relations:
$$
\frac{\partial\chi}{\partial z}=\dot{\chi},
\qquad
\frac{\partial\chi}{\partial t}=\chi',
\qquad
x_i'=c\beta_i,
\qquad
i=\overline{1,3}.
$$
Then
\begin{eqnarray}
& {\cal P}_i=\displaystyle\frac{E}{c}\beta_i,
\qquad
\int \limits_0^{z(t)}F(z)dz=\int
\limits^t_0F(z(t))c\beta_3dt,\\ \nonumber
& \dot\chi=\displaystyle\frac{E}{c^2}\frac{1}{\beta_3}\chi',
\qquad
\frac{E}{c}-({\vec{e}}_{\perp},\vec{\cal P}_{\perp})=
\frac{E}{c}(1-({\vec{\beta}}_{\perp},{\vec{e}}_{\perp})),\\ \nonumber
& z\rho_0(z)-e_3z=ct-({\vec{r}}_{\rm cl}(t),\vec{n})=F_0(t),
\qquad
z\rho_1(z)=\displaystyle\frac{c^2}{E}\rho_1(t);\\ \nonumber
& \displaystyle\left|
\frac{\partial\Phi}{\partial\vk}
\right|_{\Phi(\vk)=0}=
\frac{1}{\beta_{\parallel}}
\left\{
\psi_0+\frac{\h\omega_{cl}}{E}\frac{\rho_1(T)}T
\right\}=\frac{\psi_0}{\beta_{\parallel}}\frac{\omega^2_{\rm cl}}{\omega^2}
+O(\h^2),\\
& \omega=\omega_{\rm cl}
\displaystyle\left\{
1-\frac{\h}{2E}\frac{\omega_{\rm cl}\rho_1(T)}{\psi_0T}
\right\},
\qquad
\psi_1(t)=\omega_{\rm cl}
\displaystyle\left\{
F_0(t)+\frac{\h\omega_{cl}}{2E}F_1(t)
\right\},\\
& \omega_{\rm cl}=\displaystyle\frac{n\omega_0}{\psi_0},
\qquad
\omega_0=\frac{2\pi}T,\nonumber
\end{eqnarray}
where
\begin{eqnarray*}
& \rho_1(t)=\displaystyle\int_0^t
\{[1-({\vec{\beta}}_{\perp},{\vec{e}}_{\perp})]^2-\beta_3^2e_3^2\}
\frac{dt}{\beta_3^2},\\
& F_1(t)=\rho_1(t)-\displaystyle\frac{ \rho_1(T) }{ T\psi_0 }F_0,
\end{eqnarray*}
and for the matrix elements of transition currents
\begin{eqnarray}
& \vec{B}(n)=\displaystyle\frac{1}{\beta_{\parallel}T}
\int \limits^T_0dte^{i\psi_1}
\left\{
\vec{\beta}\left(1+\frac{ \h\omega_{\rm cl}}{2E\beta_3^2}
\left(
1-({\vec{\beta}}_{\perp},{\vec{e}}_{\perp})\right)
\right)+
\vec{k}i\frac{ \h\beta_3' }{ 2E\beta_3^{3} }
\right\},\\
& \vec{k}=(0,0,1), \qquad
\vec e_{\perp}=(e_1,e_2,0),\qquad \vec\beta_{\perp}=
(\beta_1,\beta_2,0),\\ \nonumber
& e_1=\sin\theta\cos\varphi,\qquad e_2=\sin\theta\sin\varphi,
\qquad e_3=\cos\theta.\nonumber
\end{eqnarray}

The final result for the radiation power into the
element of solid angle $d\Omega=\sin\theta\,d\theta d\varphi$ is given by
\begin{eqnarray}
& \displaystyle\frac{dW}{d\Omega}=\frac{e_0^2\beta^2_{\parallel}}{2\pi c}
\sum_{n=1}^{\infty}\frac{\omega^4}{\omega^2_{\rm cl}\psi_0}
\left(
|\alpha_\pi(n)|^2+|\alpha_\sigma(n)|^2\right),\\
& \alpha_\sigma(n)=(\vec B(n), \vec e_\varphi), \qquad
\alpha_\pi(n)=(\vec B(n),\vec e_\theta), \qquad
\psi_0=1-\beta_{\parallel}e_3,\\ \nonumber
& \vec e_\sigma=(\sin\varphi,-\sin\varphi,0),\\ \nonumber
& \vec e_\pi=(\cos\varphi\sin\theta,\sin\varphi\cos\theta,
-\sin\theta),\nonumber
\end{eqnarray}
where $c\beta_{\parallel}$ is the average drift velocity along the axis $Oz$.
The unit vector $\vec{n}=(e_1,e_2,e_3)$ points in the direction of
photon emission.

The expressions (2.1), (4.2)--(4.4) obtained above solve the problem of
including the first quantum correction into the radiation of a spinless
particle. It should be noted, that the same result can be derived by
using the solution of the Klein-Gordon equation in the field (1.1) [12].

Using the properties of conjugate Fourier series [27] in expression (4.5)
it is possible to sum over $n$ and to obtain the spectral-angular
distribution of the radiation power considering the first quantum
correction in the form:
\begin{eqnarray}
& W=\displaystyle\frac{e_0^2}{2\pi c}
\oint \frac{d\Omega}{2T}
\left[
\int \limits^t_0\frac{A^2(t)}{\psi}dt+
\frac{\h}{ET}\int\limits^t_0dt_1
\int \limits^t_0dt_2A(t_1)D(t_2)\cot{\Xi}
\right]+O(\h^2),\\
& \Xi=\pi [c(t_1-t_2)-\{
(\vec{n},\vec r_0(t_1))-(\vec{n},\vec r_0(t_2))
\}],\nonumber
\end{eqnarray}
where the double integral is understood in the sense of the
principal value, and the functions $D(t),A(t)$ have the form:
\begin{eqnarray*}
& A^2(t)=(\vec A(t),\vec e_\varphi)^2+(\vec A(t),
\vec e_\theta)^2,\qquad
\vec A(t)=[
\vec\beta(\vec n,\vec\beta')+\psi\vec\beta']/\psi^2;\\
& A(t_1)D(t_2)=(\vec A(t_1),\vec e_\sigma)
(\vec D(t_2),\vec e_\sigma)+(\vec A(t_1),\vec e_\pi)
(\vec D(t_2),\vec e_\pi);\\
& \vec D(t)=\displaystyle\frac{d}{dt}\frac{\vec B(t)}\psi,
\qquad \psi=F'_0=1-(\vec n,\vec\beta);\\
& \vec B(t)=\Big\{
\big(1-({\vec{e}}_{\perp},{\vec{\beta}}_{\perp})\big)
\beta_3\vec{A}(t)\psi^2+
[2\beta_3{\vec{\beta}}'\psi+2\beta_3\vec{\beta}(\vec{n},
\vec{\beta})-\vec{\beta}{{\beta}}_3'\psi]
\displaystyle\frac{e_3}{\beta_3^2}+\\
& +\displaystyle\frac{F_1(t)\vec g}{\psi^2}-\frac{\beta'_3}
{\beta_3^2\psi^2}\vec{k}\Big\}\frac 1{\psi^2};\\
& F_1(t)=\displaystyle\int\limits^t_0
\frac{[1-(\vec\beta_{\perp},\vec e_{\perp})]^2}
{\beta_3^2}dt-
\frac{F_0}{\omega_{\rm cl}\psi_0};\\
& \vec g=\vec\beta''\psi^2+
\vec\beta\psi(\vec{n},\vec\beta'')+
3(\vec{n},\vec\beta')[\psi\vec\beta'+
\vec\beta(\vec n,\vec\beta')].
\end{eqnarray*}

In the classical term of (4.6), one can integrate explicitly
over angles, however, in the quantum term, this integration is impossible.

Complete calculation of radiation characteristics can be done in the near-axis
approximation $\mu\to 0$ (4.1). The terms of first
order in $\mu\to 0$ in (4.2)--(4.5) imply
\begin{eqnarray*}
& \beta_3\approx\beta\approx\beta_{\parallel}, \qquad
\beta_1\approx\beta_2\approx\mu\beta_{\parallel},\\
& F_0(t)=\psi_0t-\displaystyle\int\limits^t_0
(\vec e_{\perp},\vec\beta) dt+O_\mu(\mu^2),\\
& \rho_1(t)=\beta_{\parallel}^{-2}
\left[
(1-\beta_{\parallel}^2e_3^2)t
-2\displaystyle\int\limits^t_0
(\vec e_{\perp},\vec\beta_{\perp}) dt
\right]+O_\mu(\mu^2),\\
& F_1(t)=-\beta^2\psi_0\displaystyle\int\limits^t_0
(\vec e_{\perp},\vec\beta_{\perp})dt+O_{\mu}(\mu^2).
\end{eqnarray*}

Therefore, in the near-axis approximation we have
\begin{eqnarray}
& \omega=\displaystyle\frac{ \omega_0n}{\psi_0}
\left\{
1-\frac{n}{2}\xi_0\frac{1+\beta_{\parallel}e_3}{\psi_0}
\right\}+O_{\mu}(\mu^2),\\
& \displaystyle\left|
\frac{\partial\Phi}{\partial\vk}
\right|_{\phi(\vk)=0}=\frac{\psi_0}{\beta_{\parallel}}
\left\{
1+n\xi_0\frac{ 1+\beta_{\parallel}e_3 }{\psi_0}
\right\}+O_{\mu}(\mu^2),\\
& \xi_1=n\omega_0t-\displaystyle\frac{n\omega_0}{\psi_0}
\Big(1+\frac{n}{2}\xi_0\Big)+O_{\mu}(\mu^2),\\
& \vec{B}=\displaystyle\frac{1}{\beta_{\parallel}T}
\int\limits^T_0dt \exp\{i\omega_0nt\}
\vec{\beta}\left[
1+\frac{(\vec\beta_{\perp},\vec e_{\perp})}{\psi_0}
\right]\Big(1+\frac{n\psi_0}{2}\xi_0\Big)+O_\mu(\mu^2),\nonumber
\end{eqnarray}
where $\xi_0=\h\omega/(E\beta_{\parallel}^2)$.

For the spectral-angular distribution of the radiation power, we
obtain
\begin{equation}
\frac{dW}{d\Omega}=\frac{e_0^2}{2\pi c}
\sum\limits_{n=1}^\infty n^2
[S_{\pi}(\theta,\varphi)+S_\sigma(\theta,\varphi)]
\Big[1-\frac{n\xi_0}{\psi_0}(1+2\beta_{\parallel}e_3)
\Big]+O_\mu(\mu^2)+O(\h^2),
\end{equation}
where
\begin{eqnarray*}
&& S_\pi(\theta,\varphi)=f_\pi(\theta)\left|
\displaystyle\frac{1}{T}
\int\limits^T_0(\beta_1\sin{\varphi}-\beta_2\cos{\varphi})
e^{i\omega_0nt}dt \right|^2,\\
&& S_\sigma(\theta,\varphi)=f_\sigma(\theta)\left|
\displaystyle\frac{1}{T}
\int\limits^T_0(\beta_1\cos\varphi-\beta_2\sin\varphi)
e^{i\omega_0n t}dt\right|^2,\\
&& f_\sigma(\theta)=\displaystyle\frac 1{\psi_0^3},\qquad
f_\pi(\theta)=\frac{(\beta_{\parallel}-\cos\theta)^2}{\psi_0^5}.
\end{eqnarray*}

After integration over angles, we find the spectral
distribution of radiated power in the following form:
\begin{eqnarray}
&& W=\displaystyle\frac{e_0^2\omega_0^2}{c}
\sum\limits_{n=1}^\infty n^2(S_\pi+S_\sigma)
\sum_{k=1}^2
\left| \frac{1}{T}
\int\limits^T_0e^{i\omega_0nt}\beta_kdt
\right|^2+O_{\mu}(\mu^2)+O(\h^2)=\\ \nonumber
&&\qquad =\displaystyle\frac{e_0^2\omega_0^2\beta^2_{\parallel}}{c}
\sum\limits_{n=1}^\infty n^2(S_\pi+S_\sigma)
\sum_{k=1}^2
\left| \frac 1l \int\limits^l_0 H_k(z) e^{i\omega_nz}dz
\right|^2+O_{\mu}(\mu^2)+O(\h^2),\\
&& S_\pi=\displaystyle\frac 13\frac 1{(1-\beta_{\parallel}^2)^2}
\left[
1-\frac 15 \xi_0n\frac{5+19\beta^2_{\parallel}}{1-\beta^2_{\parallel}}
\right],\\ \nonumber
&& S_\sigma=\displaystyle\frac{1}{(1-\beta_{\parallel}^2)^2}
\left[
1-\xi_0 n\frac{1+3\beta_{\parallel}^2}{1-\beta^2_{\parallel}}
\right],\nonumber
\end{eqnarray}
where $\vec{H}(z)=(-\dot{\cal A}_2(z),\dot{\cal A}_1(z),0)$ is the
magnetic field and $\omega_n=2\pi n/l$.

For summation over the spectrum we use the Parseval equality and
the properties of conjugate Fourier series [27] were used:
\begin{eqnarray*}
&&\displaystyle\sum\limits_{l=1}^2\sum\limits_{n=1}^{\infty}n^2
\left|
\frac{1}{T}\int\limits^T_0e^{i\omega_0nt}\beta_ldt\right|^2=
\frac{1}{\omega_0^2}\frac{1}{2T}\int\limits^T_0
(\vec\beta'_{\perp})^2dt=\frac{1}{2\omega_0^2}\delta_0,\\
&&\displaystyle\sum\limits_{l=1}^2\sum\limits_{n=1}^{\infty}n^{3}
\left|
\frac{1}{T}\int_0^Te^{i\omega_0nt}\beta_{l}dt
\right|^2=\\
&&\qquad\qquad=\displaystyle\frac{1}{\omega_0^{3}}\frac{1}{2T^2}
\int\limits^T_0dt\int\limits^T_0d\tau
(\vec\beta'_{\perp}(t),\vec\beta''_{\perp}(\tau))
\cot{(\frac{t-\tau}{T}\pi)}=\frac{1}{2\omega_0^3}\delta_1.
\end{eqnarray*}

The last integral is understood in the sense of the principle value.
So, the total radiation power with account of the first quantum correction is
\begin{eqnarray}
& W=W_{\rm cl}(I_{\pi}+I_{\sigma}),\\
& I_\pi=\displaystyle\frac{1}{4}-\frac{\h}{E\beta_{\parallel}^2}
\frac{\delta_1}{\delta_0}\frac{5+19\beta_{\parallel}^2}
{20(1-\beta_{\parallel}^2)},\\ \nonumber
& I_\sigma=\displaystyle\frac{3}{4}-\frac{\h}{E\beta_{\parallel}}
\frac{\delta_1}{\delta_0}
\frac{3+9\beta_{\parallel}^2}{4(1-\beta_{\parallel}^2)},\\ \nonumber
& W_{\rm cl}=\displaystyle\frac{2}{3}\frac{e_0^2}{c(1-\beta_{\parallel}^2)^2}
\frac{1}{T}
\int\limits^T_0dt(\vec\beta_{\perp}')^2
=\frac{2}{3}\frac{e_0^4\beta_{\parallel}^2}
{m_0^2c^{3}(1-\beta_{\parallel}^2)}
\frac{1}{l}
\int\limits^l_0(\vec{H}(z))^2dz.\nonumber
\end{eqnarray}
If a component of the magnetic field is equal to zero, then we have
the radiation power in the plane undulator regime [9, 10].

{\bf Remark}. From a ratio (4.11) it follows that in on-axis approximation
the maximum in distribution on harmonics it is necessary that number $n$ for
which factor $H(n)$ of expansion in a Fouier series for a magnetic field
force is maximum. In first, on visible, this conclusion verifying the
qualitative Motz results [28] (in the respect that not without fail maximum
on the module is first ($n=1$) Fouier factor) was made in work [29].
In particular, from (4.11) it follows that of magnetic fields, for which
the influence of a harmonic $H(n_{\rm cr})$ is essential, where
$n_{\rm cr}\sim (1-\beta^2_{\parallel})/\xi_0$, the semiclassical
expansion of (1.9) type is not true, as the quantum amendment becomes
comparable with classical summand.

\section{The radiation power in the helical undulator}

In general, it is impossible to integrate the spectral angular distribution
over all angles and to sum it over the whole spectrum. The total radiation
power including the first quantum correction can be obtained in the
ultrarelativistic case only for special magnetic field, in which the
electron moves along the helical trajectory. Let us consider
the periodical magnetic field
\begin{equation}
\vec H=\{-H_0\cos az,-H_0\sin az,0\}
\end{equation}
with the period $l=2\pi/a$. The relativistic equation of motion is
\begin{eqnarray}
& x''=z'\omega_0\sin{az},\qquad
y''=-z'\omega_0\cos{az}, \\ \nonumber
& z''=\omega_0(x'\sin{az}+y'\cos{az}),\\
& \omega_0=e_0cH_0/E, \qquad e_0=|e|.\nonumber
\end{eqnarray}

In the general case (with the arbitrary initial conditions), the solution of
system (5.2) is expressed by the elliptical integral.
However, in the case of interest, when the condition of absence of the
average drift along the axis $Ox$ and $Oy$ is fulfilled and the projection
of initial momentum on this axis is zero, the dependence of
coordinates on time is of the form
\begin{equation}
\vec{r}(t)=\{R\sin{\omega_0t},-R\cos{\omega_0t},c\beta_{\parallel}t\}.
\end{equation}

The system (5.3) describes the motion of a particle along the helix with
frequency of spiral motion $\omega_0=2\pi/T=c\beta_{\parallel}a$
and the radius of spiral $R=\beta_{\perp}/(\beta_{\parallel}a)$,
$\beta_{\perp}=e_0H/(Ea)$.

In the field (5.1), the spectral-angular distribution of radiation power
is defined by the general formulae (4.4) and (4.5) and is of the form (see
Appendix A1):
\begin{eqnarray}
& W=\displaystyle\frac{e_0^2}{c}\omega_0^2\sum\limits_{n=1}^{\infty}
\int\limits_0^{\pi}\frac{\sin\theta\,d\theta}
{(1-\beta_{\parallel}\cos\theta)^3}
[|\alpha_\pi(n)|^2+|\alpha_\sigma(n)|^2],\\
& |\alpha_\pi(n)|^2=\left(\displaystyle\frac{\cos{\theta}-\beta_{\parallel}}
{\sin{\theta}}\right)^2
\left\{
{\cal J}_n^2(z)-
\frac{\h\omega_0n}{E\beta_{\parallel}^2}
\Big[
{\cal J}_n^2(z)\Big(\frac 32
\frac{1+\beta_{\parallel}\cos{\theta}}{1-\beta_{\parallel}\cos
\theta}+
\frac{z^2}{n^2}
\Big)-
\right.\\ \nonumber
&\left.
-\displaystyle\frac{z}{2}\left(1-\frac{z^2}{n^2}\right)
\dot{\cal J}_n(z){\cal J}_n(z)\Big]\right\},\\ \nonumber
& |\alpha_\sigma(n)|^2=\beta_{\perp}
\left\{
\dot{\cal J}^2_n(z)-
\displaystyle\frac{\h\omega_0n}
{E\beta_{\parallel}^2}
\Big[\dot{\cal J}^2_n(z)
\left(
\frac{1+2\beta_{\parallel}\cos\theta}{1-\beta_{\parallel}
\cos\theta}+
\frac{z^2}{n^2}
\right)-\right.\\ \nonumber
& \left.
-\displaystyle\frac{z}{2}\Big(1-\frac{z^2}{n^2}\Big)
\dot{\cal J}_n(z)\ddot{\cal J}_n(z)\Big]\right\},\nonumber
\end{eqnarray}
where
$$
z=\frac{n\beta_{\perp}\sin\theta}{1-\beta_{\parallel}\cos\theta},
\quad
\dot{\cal J}_n(z)=\frac{d}{dz}{\cal J}_n(z),
$$
and ${\cal J}_n(z)$ are the Bessel functions [30]. The expression (5.4) as
$\h\to 0$ coincides with the well-known formula for radiation
power for the motion along the helix in a homogeneous magnetic field [21].

The integration over angles and the summation over all frequencies in
the quantum term of expression (5.4) can be done only in the ultrarelativistic
limit $\beta\to 1$, $\beta^2=\beta_{\parallel}^2+\beta_{\perp}^2$.
We change to the system of reference, which moves along the initial
axis $Oz$ with velocity $\beta_{\parallel}$ [21], and we use the
asymptotic representation of the Bessel function ${\cal J}_n(z)$ and its
derivatives as the argument $z\to\nu-0$ [21]. The result for the total
radiation power with account of the first quantum corrections is
\begin{eqnarray}
& W=W_{\rm cl}(f_\pi+f_\sigma)=W_{\rm cl}
\left(1-\displaystyle\frac{55\sqrt{3}}{16}\frac{\h}{m_0cR}
\left(\frac{E}{m_0c^2}\right)^2
(1-\beta_{\parallel}^2)
\right),\\
& W_{\rm cl}=\displaystyle\frac{2}{3}\frac{e_0^2c}{R^2}
\left(\frac{E}{m_0c^2}\right)^4,\nonumber
\end{eqnarray}
and the $f_\pi$, $f_\sigma$ are $\pi$- and $\sigma$-components
of polarization which are equal to:
\begin{eqnarray*}
&& f_\pi=\displaystyle\frac{1}{8}-\frac{5\sqrt{3}}{16}\frac{\h}{m_0cR}
\left(\frac{E}{m_0c^2}\right)^2
(1-\beta_{\parallel}^2),\\
&& f_\sigma=\displaystyle\frac{7}{8}-\frac{50\sqrt{3}}{16}\frac{\h}{m_0cR}
\left(\frac{E}{m_0c^2}\right)^2
(1-\beta_{\parallel}^2).
\end{eqnarray*}

\section{The probability of transition with the spin flip}

The probability of transition $w(\z,\z')$ with the spin
flip can be found by analogy with the expression (2.5):
\begin{eqnarray}
& w(\z,\z')=\displaystyle\frac{e_0^2}{2\pi\h}
\int\limits_0^{2\pi}d\varphi\int\limits_0^{\pi}
\frac{\sin\theta\,d\theta}{\psi_0}
\sum\limits_{n=[\nu_0]}^{\infty}\vk
\{|\alpha_\pi^{\uparrow\downarrow}(n,\z)|^2+
|\alpha_\sigma^{\uparrow\downarrow}(n,\z)|^2\},\\
& \alpha_\pi^{\uparrow\downarrow}(n,\z)=(\vec e_\pi,\vec B^{\uparrow
\downarrow}(n,\z)),qquad
\alpha_\sigma^{\uparrow\downarrow}(n,\z)=(\vec e_\sigma,
\vec B^{\uparrow\downarrow}(n,\z)),\nonumber
\end{eqnarray}
where $\nu_0$ is determined by the condition $\vk(\nu_0)=0$. In matrix
elements $B^{\uparrow\downarrow}(n,\z)$ it is convenient to change the
integration variables $dz=(c^2p/E)dt$. The result is
\begin{eqnarray}
&\left\{
\begin{array}{c}
B_1^{\uparrow\downarrow}(n,\z)\\B_2^{\uparrow\downarrow}(n,\z)\\
B_3^{\uparrow\downarrow}(n,\z)\end{array}\right\}
=\displaystyle\frac{\h\omega m_0c^2}{2E^2\beta^2}
\frac{1}{\beta_{\parallel}T}
\int\limits^T_0
\frac{e^{i\psi_2(t)}dt}{(\beta_1^2+\beta_2^2)^{1/2}}
\left\{
\begin{array}{c}
i\beta\beta_2-\z\beta_3\beta_1\\
-i\beta\beta_1-\z\beta_3\beta_2\\
\z({\vec{\beta}}_{\perp},{\vec{\beta}}_{\perp})\end{array}
\right\},\\
& \psi_2(t)=\omega F_0(t)+\z\rho_3(t),\qquad
F_0(t)=\displaystyle\frac{1}{c}[ct-(\vec r_{\rm cl}(t),\vec{n})],\\ \nonumber
&\rho_3(t)=\beta\displaystyle\int\limits^t_0
\frac{\beta_2'\beta_1-\beta_1'\beta_2}
{\beta_3(\beta_1^2+\beta_2^2)}
dt,\qquad
\beta_i'=\frac{d\beta_i}{dt},\\ \nonumber
& \beta^2=(\vec\beta,\vec\beta),\qquad
c\vec\beta=\displaystyle\frac{d\vec r_{\rm cl(t)}}{dt},\qquad
\vec\beta_{\perp}=(\beta_1,\beta_2,0),\\ \nonumber
& \omega=\displaystyle\frac{1}{\omega_0}
\left(
\omega_0n-\z\frac{\rho_3(T)}{T}
\right),\qquad
\omega_0=\frac{2\pi}{T}.\nonumber
\end{eqnarray}

Now let us consider the model of a helical undulator (5.1), which is
characterized by a helical electron trajectory:
\begin{equation}
\vec r_{\rm cl}(t)=\frac c{\omega_0}\{\beta_{\perp}\sin\omega_0t,
-\beta_{\perp}\cos\omega_0t,\omega_0\beta_{\parallel}t\}.
\end{equation}

In this case
\begin{eqnarray*}
& \psi_2(t)=n\omega_0t-z\sin(\varphi-\omega_0t),\qquad
\rho_3(t)=\displaystyle\frac{\beta}{\beta_{\parallel}}\omega_0t;\\
& \beta_{\perp}=\displaystyle\frac{e_0H}{Ea}=\frac{e_0Hl}{2\pi E}=
\frac{e_0Hl}{2\pi m_0c^2\gamma};\\
& z=z_0\left(n-\displaystyle\frac{\z\beta}{\beta_{\parallel}}\right),\qquad
z_0=\frac{\beta_{\perp}\cos\theta}{\psi_0}
\end{eqnarray*}
and it is possible to integrate over the time [13]
\begin{eqnarray}
&& \alpha_\pi(n,\z)=i\displaystyle\frac{\h\omega}{2E\beta^2}(1-\beta^2)^{1/2}
e^{-in\varphi}
\left\{
\z\beta_{\parallel}\dot{\cal J}_n(z)+\beta\frac{n}{z}{\cal J}_n(z)
\right\},\\
&& \alpha_\sigma(n,\z)=-\displaystyle\frac{\h\omega}{2E\beta^2}
(1-\beta^2)^{1/2}e^{-in\varphi}
\left\{
\left(
\beta\dot{\cal J}_{n}(z)+
\z\beta_{\parallel}\frac nz {\cal J}_n(z)\right)
\cos{\theta}+
+\z\beta_{\perp}{\cal J}_n(z)\sin\theta\right\}.
\end{eqnarray}

Integration of the expressions over angles and their
summation over the spectrum is possible in the case of the near-axis
approximation, which is characterized by the maximal deflection angle
of electron velocity from the undulator axis
$\mu=\beta_{\perp}/\beta_{\parallel}$ (4.1).
In the case of helical undulator this means that the parameter
$z_0=(\beta_{\perp}\sin\theta)/\psi_0^2$ is very small. Keeping in
(6.4), (6.5) only terms, which are liniar over $z_0\ll 1$, we obtain
\begin{eqnarray*}
& \beta\sim\beta_3\sim\beta_{\parallel},\qquad
\omega=\omega_0(n-\z)/\psi_0;\\
& \alpha_\pi(n,\z)=i\displaystyle\frac{\h\omega}{2E\beta_{\parallel}^2}
(1-\beta^2)^{1/2}e^{-in\varphi}
\left\{
\frac{1}{2}\beta(1-\z)\delta_{n,-1}-\right.\\
& \qquad\left.
-\displaystyle\frac{1}{2}\z\beta_{\parallel}z\delta_{n,0}+\frac{1}{2}
\beta(1+\z)\delta_{n,1}+\frac{1}{4}\beta(1-\z)z\delta_{n,2}\right\};\\
& \alpha_\sigma(n,\z)=-\displaystyle\frac{\h\omega}{4E\beta_{\parallel}^2}
(1-\beta^2)^{1/2}e^{-in\varphi}
\Big\{
-\left(
\beta(1+\z)\cos\theta+\z\beta_{\perp}z\sin\theta
\right)\delta_{n,-1}-\\
&-(z\beta\cos\theta-2\z\beta_{\perp}\sin\theta)\delta_{n,0}+
[\beta(1+\z)\cos\theta+\z z\beta_{\perp}\sin\theta]\delta_{n,1}+
\displaystyle\frac{z}{2}\beta(1+\z)\cos\theta\,\delta_{n,2}\Big\}.
\end{eqnarray*}
Here, we used the following asymptotic representation of Bessel functions
[30]:
$$
{\cal J}_n(z)
\mathop{\approx}\limits_{z\to 0}
\frac{z^n}{2^nn!}, \qquad
{\cal J}_{-n}(z)=(-1)^n{\cal J}_n(z).
$$

After summation over the spectrum we have
\begin{eqnarray}
& w(\z,-\z)=\displaystyle\frac{\h e_0^2\omega_0^3}
{16cE^2\beta^4}\beta_{\perp}^2\{f_\pi+f_\sigma\},\\
& f_\sigma=\displaystyle\int\limits_0^\pi\frac{\beta^2(1-\beta^2)}{\psi_0^6}
\sin\theta\,d\theta,\\ \nonumber
& f_\pi=\displaystyle\int\limits_0^\pi\frac{(\psi_0-\z)^2(1-\beta^2)}
{\psi_0^6}\sin^3\theta\, d\theta.\nonumber
\end{eqnarray}

Finally, we have:
\begin{eqnarray}
& w(\z,-\z)=\displaystyle\frac{\h\omega_0^3\beta^2_{\perp}e_0^2}
{30E^2\beta^4c}
\frac{5-2\beta^2+3\beta^4}{(1-\beta^2)^3}
\left\{
1-\z\frac{5-5\beta^2}{5-2\beta^2+3\beta^4}
\right\},\\
& w(\z,-\z)=\displaystyle\frac{1}{2\tau}\{1-\z\Gamma(\beta)\},\\ \nonumber
& \tau=\displaystyle\frac{15E^2\beta^4(1-\beta^2)^3c}
{\h\omega_0^{3}\beta_{\perp}^2e_0^2(5-2\beta^2+3\beta^4)}.\nonumber
\end{eqnarray}
The function
$$
\Gamma(\beta)=\frac{5-6\beta^2}{5-2\beta^2+3\beta^4}
$$
is a monotonic function within the limits:
\begin{equation}
1=\Gamma(0)\geq\Gamma(\beta)\geq\Gamma(1)=0.
\end{equation}
It follows from (6.6) that the probability of the electron radiation with
spin flip is a function of the initial orientation of the
electron spin. This fact causes the effect of radiational
selfpolarization of spin in a bunch of electrons [21, 22]. This means that,
under certain condition
\begin{equation}
\frac{\beta_{\perp}}{\beta(1-\beta^2)^{1/2}}\ll
1,\qquad
\beta_{\perp}=\frac{e_0H}{El},\qquad
\beta^2=1-(\frac{m_0c^2}{E})^2
\end{equation}
independently of the initial orientation of the spin of particles in a
bunch, the primary spin orientation along the electrons motion
will be determined. It follows from (6.7)  that the polarization is
absent, as $\beta\to 1$, but it is complete as $\beta\to 0$. However,
the condition (6.9) does not allow us to consider these limiting cases.

Let us consider in (6.6) the relativistic particles with $\beta\to 1$.
Then, we find the relaxation time $\tau$ and asymptotic degree of
polarization $p$:
\begin{equation}
\tau=\frac{5}{2}\frac{m^2c^5}{\omega_0^3\beta_{\perp}^2e_0\h}
(1-\beta^2)^2,\;\;\;p=\frac{5}{6}(1-\beta^2),\qquad
\beta^2=1-\gamma^2.
\end{equation}

These expressions within an accuracy up to $O(\gamma^{-6}),
\gamma\gg 1$ coincide with expressions previously obtained in [7].

\section*{Acknowledgments}
This work was partially supported by Russian Foundation for Basic Research
(Grants 96-01-04578 and 97-02-162179). MMN was supported by
International Science Foundation (Grant 539-p). AYuT, whose work is carried out within the
research program of International Center for Fundamental Physics in
Moscow, is a fellow of INTAS Grant 93-2492.

\section*{Appendix. The first quantum correction of the boson
radiation in a helical undulator}

After the summation over the spin, the spectral-angular distribution of the
radiation intensity of the electron moving in field (5.1) with
the first quantum correction has the form
$$
\begin{array}{c}
W=\displaystyle\frac{e_0^2}{c}\omega_0^2
\sum^{\infty}_1n^2
\int\limits_0^{\pi}
\frac{\sin{\theta}d\theta}{(1-\beta_0\cos{\theta})^{3})}
\left(1-\frac{\h\omega_0}{E\beta_0^2}n
(2+\frac{z^2}{n^2})\right)
(|\alpha_\pi(n)|^2+|\alpha_\sigma(n)|^2),\\
\omega_0=\displaystyle\frac{2\pi}{T},\;\;\;
z=\frac{n\beta_{\perp}\sin{\theta}}
{1-\beta_0\cos{\theta}},
\end{array}
\eqno{(A.1)}$$
where
$$
|\alpha_\pi(n)|^2=|B_1(n)\cos\theta-B_3(n)\sin\theta
|^2,\qquad |\alpha_\sigma(n)|^2=|B_2(n)|^2;
$$
$$
\begin{array}{c}
\vec B(n)=\displaystyle\frac{1}{2\pi}\int\limits_0^{2\pi}d\eta
\exp\left\{i\left(n\eta-z_1\sin{\theta}-
\frac{\h\omega_0n}{E\beta_0^2(1-\beta_0\cos{\theta})}
z_2\sin{2\eta}\right)\right\}\times\\
\times
\vec{\beta}\left\{1-\displaystyle\frac{\h\omega_0}{2E\beta_0^2}
\frac{n}{1-\beta_0\cos{\theta}}(1-\beta_{\perp}\sin{\theta}\cos{\theta})
\right\};\\
z_1=z\Big(1-\displaystyle\frac{\h\omega_0n}{2E\beta_0^2}\Big(1+\frac{1}{2}
\frac{\xi^2}{n^2}\Big)\Big),
\qquad
z_2=z\frac{1}{8}\beta_{\perp}\sin\theta.
\end{array}
\eqno{(A.2)}$$

We will integrate (A.2) over $\eta$ using the following relations:
$$
\begin{array}{l}
\displaystyle\frac{1}{2\pi}\int\limits_0^{2\pi}
e^{i(n\eta-\xi\sin\eta)}d\eta={\cal J}_n(\xi);\\
\displaystyle\frac{1}{2\pi}\int\limits_0^{2\pi}
e^{i(n\eta-\xi\sin\eta)}\sin\eta\,d\eta=
i\dot{\cal J}_n(\xi);\\
\displaystyle\frac{1}{2\pi}\int\limits_0^{2\pi}
e^{i(n\eta-\xi\sin\eta)}\cos\eta\,d\eta
=\frac{n}{\xi}{\cal J}_n(\xi);\\
\displaystyle\frac{1}{2\pi}\int\limits_0^{2\pi}
e^{i(n\eta-\xi\sin\eta)}\sin^2\eta\,d\eta
=-\ddot{\cal J}_n(\xi);\\
\displaystyle\frac{1}{2\pi}\int\limits_0^{2\pi}
e^{i(n\eta-\xi\sin\eta)}\cos^2\eta\,d\eta
=\frac{n^2}{\xi^2}{\cal J}_n(\xi)-
\frac{1}{\xi}\dot{\cal J}_n(\xi);\\
\displaystyle\frac{1}{2\pi}\int\limits_0^{2\pi}
e^{i(n\eta-\xi\sin\eta)}\sin{2\eta}\,d\eta
=i\frac{2n}{\xi}\left(\dot{\cal J}_n(\xi)-
\frac{1}{\xi}{\cal J}_n(\xi)\right);\\
\displaystyle\frac{1}{2\pi}\int\limits_0^{2\pi}
e^{i(n\eta-\xi\sin\eta)}\sin\eta\cos\eta\,d\eta
=i\frac{n}{\xi}\left(\dot{\cal J}_n(\xi)-
\frac{n}{\xi}{\cal J}_n(\xi)\right);\\
\displaystyle\frac{1}{2\pi}\int\limits_0^{2\pi}
e^{i(n\eta-\xi\sin\eta)}\sin{2\eta}\sin\eta\,d\eta
=2i\left\{-\frac{2n^2}{\xi^2}{\cal J}_n(\xi)+\frac{n^2+1}{\xi^2}
\dot{\cal J}_n(\xi)-\frac{1}{\xi}\ddot{\cal J}_n(\xi)\right\};\\
\displaystyle\frac{1}{2\pi}\int\limits_0^{2\pi}
e^{i(n\eta-\xi\sin\eta)}\sin{2\eta}\sin\eta\,d\eta
=-2n\left\{\frac{1}{\xi}\ddot{\cal J}_n(\xi)-\frac{2}{\xi^2}
\dot{\cal J}_n(\xi)+\frac{2}{\xi^2}{\cal J}_n(\xi)\right\}
\end{array}
\eqno{(A.3)}
$$
and
$$
\vec{\beta}=\left\{\beta_{\perp}\cos{\eta},-\beta_{\perp}\sin{\eta},
\beta_0\right\}.
$$

As a result for the matrix elements we obtain
$$
\begin{array}{c}
B_1(n)=\beta_{\perp}\left\{\displaystyle\frac{n}{z_1}{\cal J}_n(z_1)
\left(1-\frac{\h\omega_0}{2E\beta^2_0}
\frac{n}{1-\beta_0\cos\theta}\right)
+\frac{\h\omega_0}{E\beta^2_0}
\frac{n}{1-\beta_0\cos\theta}\right.\times\\
\times
\left[\displaystyle\frac{1}{2}\beta_{\perp}\sin\theta
\left(\frac{n^2}{z^2}{\cal J}_n(z_1)-\frac{1}{z}
\dot{\cal J}_n(z_1)\right)+\right.\\
\left.\left.
+2z_2\left(\displaystyle\frac{-2n^2}{z^2}{\cal J}_n(z_1)+
\frac{n^2+1}{z^2}\dot{\cal J}_n(z_1)-
\frac{1}{z}\ddot{\cal J}_n(z_1)\right)\right]\right\};\\
B_2(n)=-i\beta_{\perp}\left\{\left(1-\displaystyle\frac{\h\omega_0}
{2E\beta_0^2}
\frac{n}{1-\beta_0\cos\theta}\right)\dot{\cal J}_n(z_1)+
\frac{\h\omega_0}{E\beta_0^2}\frac{n}{1-\beta_0\cos\theta}\right.\times\\
\times
\left[\displaystyle\frac{n}{2z}\beta_{\perp}\sin\theta\left(
\dot{\cal J}_n(z_1)-\frac{1}{z}{\cal J}_n(z_1)\right)+\right.\\
\left.\left.
+2z_2\left(\displaystyle\frac{n}{z}\ddot{\cal J}_n(z_1)-
\frac{2n}{z^2}\dot{\cal J}_n(z_1)+
\frac{2n}{z^2}{\cal J}_n(z_1)\right)\right]\right\};\\
B_3(n)=\beta_0\left\{\left(1-\displaystyle\frac{\h\omega_0}{2E\beta_0^2}
\frac{n}{1-\beta_0\cos\theta}\right){\cal J}_n(z_1)+
\frac{\h\omega_0}{E\beta_0^2}\frac{n}{1-\beta_0\cos\theta}\right.\times\\
\times\left.
\left[\displaystyle\frac{n}{2z}\beta_{\perp}\sin\theta
{\cal J}_n(z_1)+2z_2\left(\frac{n}{z}\dot{\cal J}_n
(z_1)-\frac{n}{z^2}{\cal J}_n(z_1)\right)\right]\right\}.
\end{array}
\eqno{(A.4)}
$$

Substituting (A.4) to (A.1) we obtain (5.4).

\end{document}